\documentstyle[12pt]{article}

\begin{document}
\thispagestyle{empty}

\begin{center}
\LARGE \tt \bf{Chern-Simons electrodynamics in (2+1)-space-times with torsion. }
\vspace{0.5cm}
\end{center}

\vspace{1cm}

\begin{center} {\large L.C. Garcia de Andrade\footnote{Departamento de
F\'{\i}sica Te\'{o}rica - Instituto de F\'{\i}sica - UERJ

Rua S\~{a}o Fco. Xavier 524, Rio de Janeiro, RJ

Maracan\~{a}, CEP:20550-003 , Brasil.

E-Mail.: garcia@symbcomp.uerj.br}}
\end{center}

\vspace{1cm}

\begin{abstract}
Chern-Simons electrodynamics in 2+1-spacetimes with torsion is investigated.
We start from the usual Chern-Simons (CS) electrodynamics Lagrangian and Cartan torsion is introduced in the covariant derivative and by a direct coupling of torsion vector to the CS field. 
Variation of the Lagrangian with respect to torsion shows that Chern-Simons field is proportional to the product of the square of the scalar field and torsion.
The electric field is proportional to torsion vector and the magnetic flux is computed in terms of the time-component of the two dimensional torsion.
Contrary to early massive electrodynamics in the present model the photon mass does not depend on torsion.
\end{abstract}

\vspace{1.0cm}       
\begin{center}
\Large{PACS number(s) : 04.20.Dw}
\end{center}

\newpage
\pagestyle{myheadings}
\markright{\underline{Chern-Simons electrodynamics in (2+1)-space-times with torsion.}}

Recently Antilon,Escalona and Torres (AET) \cite{1} investigate the structure and properties of of vortices and domain walls in a Chern-Simons electrodynamics in 2+1-spacetimes without torsion.
Their model is described by a gauge field interacting witha complex scalar fields including two parity time-violating terms,namely the Chern-Simons and the anomalous magnetic terms.
Self-dual relativistic vortices are presented and discussed in detail in their paper.
A Maxwell-Chern-Simons gauge theory can be constructed if a magnetic moment interaction is added between the scalar and the gauge fields.
Earlier I have investigated domain walls with torsion \cite{2,3} in four-dimensional non-Riemannian space-time.
I have also during some years to work out a model in 4-D spacetimes with torsion for a massive electrodynamics.
Despite of the fact that the breaking of gauge symmetry introduced by torsion generated a gauge problem for this theory \cite{4} Sivaram and myself were able to place some cosmological \cite{5} and astrophysical limits \cite{6} on the photon mass which agree with experimental bounds.
These results in fact could point to an indirect evidence for torsion.
All these works led us to the present paper where we investigate the extension of the CS theory where we obtain a result for the photon mass which this time does not depend on torsion.
This in no way represents an agreement with Hehl's argument \cite{7} that torsion does not interact with torsion even because by variation of the Lagrangian we obtain the constraint equation that torsion is algebraically related with the electromagnetic CS field.
The magnetic flux depends on torsion vector time component and the scalar field squared.
The electric field is also obtained in terms of the spatial part of the torsion vector.
It is interesting to note that in our model the CS field is proportional to torsion in the same way the CS field in the AET model the CS is related to the electromagnetic vector potential.
It also reminds Hammond's electrodynamics with torsion where the torsion vector is proportional to the electromagnetic vector potential.
Since in the AET model the CS field is proportional to the vector current and here it is proportional to torsion vector we could say that here torsion represents a vectorial source for this electrodynamics.
Let us now consider the Lagrangian of our model as
\begin{equation}
L = - \frac{1}{4} {F}_{\lambda}{F}^{\lambda}+ \frac{k}{4} {F}^{\lambda} {A}_{\lambda}+ \frac{f}{4}F^{\lambda} {S}_{\lambda}+ \frac{1}{2}{|{D}_{\lambda}{\phi}|}^{2}   
\label{1}
\end{equation}
Greek letters represent indices that go from zero to two.
Here $ {F}_{\lambda}={\epsilon}_{\lambda\alpha\beta}{F}^{\alpha\beta}$ is the CS field and ${F}^{\alpha\beta}$ is the electromagnetic field tensor,$ {S}^{\lambda}$ is the torsion vector and ${D}_{\lambda} $ is the covariant derivative
\begin{equation}
D_{\beta}={{\partial}}_{\beta}-ie{A}_{\beta}-ig{\epsilon}_{\beta\alpha\gamma}F^{\alpha\gamma}-if'S_{\beta}
\label{2}
\end{equation}
Variation of Lagrangian (\ref{1}) with respect to torsion vector yields
\begin{equation}
F^{\beta}=f_{0}{\phi}^{2}S^{\beta}
\label{3}
\end{equation}
the relation between the CS field and torsion appear naturally in terms of the two dimensional electric field E and the scalar magnetic field B as    
\begin{equation}
E=f_{0}{\phi}^{2}S 
\label{4}
\end{equation}
and 
\begin{equation}
B=f_{0}{\phi}^{2}S_{0}
\label{5}
\end{equation}
The magnetic flux is then given by
\begin{equation}
{\Phi}=f_{0}{\int}{\phi}^{2}S_{0}d^{2}x
\label{6}
\end{equation}
Substitution of relation (\ref{3}) into Lagrangian (\ref{1}) yields the new Lagrangian
\begin{equation}
L'=-\frac{1}{4}F_{\lambda}F^{\lambda}+\frac{k}{4}F^{\lambda}A_{\lambda}+\frac{f}{4}F^{\lambda}S_{\lambda}+\frac{1}{2}{\partial}_{\beta}
{\phi}{\partial}^{\beta}{\phi}-e^{2}A_{\beta}A^{\beta}{\phi}^{2}-
\frac{1}{2}g^{2}F_{\beta}F^{\beta}{\phi}^{2}-\frac{{f'}^{2}}{{f_{0}}^{2}}S_{\beta}S^{\beta}{\phi}^{2}
\label{7}
\end{equation}
Variation now of the last Lagrangian in terms of the electromagnetic
vector potential yields
\begin{equation}
\frac{1}{4}(1+g^{2}{\phi}^{2}){\epsilon}^{\beta\alpha\gamma}
{\partial}_{\beta}F_{\gamma}+\frac{k}{4}F^{\alpha}+e^{2}
{\phi}^{2}A^{\alpha}=0
\label{8}
\end{equation}
One may notice from this formula that it has a Maxwell-Proca term $(m^{2})_{\gamma}=e^{2}{\phi}^{2}$,here our present model differs to the models of massive electrodynamics models we worked out with C.Sivaram \cite{5} where the square of the photon mass is proportional to the square of torsion.Notice also that when the photon mass is constant the scalar field is also constant and the theory reduces to the CS theory with torsion where as we shall see bellow a massive torsion replaces a massive photon.Let us finally perform the variation of the last Lagrangian with respect to the scalar field ${\phi}$ we obtain the following Klein-Gordon-like equation 
\begin{equation}
{\nabla}^{2}{\phi}+(e^{2}A^{2}+g^{2}F^{2}+{f'}^{2}S^{2}){\phi}=0
\label{9}
\end{equation}
The coefficient of ${\phi}$ in this last equation represents a torsion mass which appears very often in some formulations of Quantum gravity in spacetimes with torsion \cite{8}.Let us finally consider some applications in the case of time independent space-time. In this case the CS-Cartan field equations reduce to
\begin{equation}
\frac{1}{4}(1+g^{2}{\phi}^{2}){\nabla}.E+\frac{k}{4}B+e^{2}{\phi}^{2}A^{0}=0
\label{10}
\end{equation}
which can be rewriten as follows
\begin{equation}
\frac{1}{4}(1+g^{2}{\phi}^{2}){\nabla}^{2}A^{0}+kB+e^{2}{\phi}^{2}A^{0}=0
\label{11}
\end{equation}
The last equation can be rewriten in the form
\begin{equation}
A^{0}=-\frac{4kB}{(1+4e^{2}{\phi}^{2}A^{0}+g^{2}{\phi}^{2}
{\nabla}^{2}A^{0})}
\label{12}
\end{equation}
Notice that in this case the electromagnetic potential vanish when the magnetic field vanishes.
\section*{Acknowledgements}
\hspace{0.6cm}I would like to thank Universidade do Estado do Rio de Janeiro
for financial Support.


\begin{thebibliography}{8}
\bibitem{1}A.Antillon J.Escalona and M.Torres,Phys.Rev.D 55,(1997),10,6327.
\bibitem{2}L.C.Garcia de Andrade,Gen.Rel.and Grav.J.(1998)30,11,1629.
\bibitem{3}L.C.Garcia de Andrade,J.Math.Phys.(1998),1,352.
\bibitem{4}V.de Sabbata and C.Sivaram,Spin and Torsion in Gravitation,(1995)
World Scientific.
\bibitem{5}L.C.Garcia de Andrade and C.Sivaram,Astrophysics and 
Space Science,209:(1993)109.
\bibitem{6}C.Sivaram and L.C.Garcia de Andrade,Torsion and Plasma 
Screening in Astrophysics,(1994),DFT-UERJ internal reports.
\bibitem{7}F.W.Hehl,Found,Phys.(1985)415.
\bibitem{8}I.Buchbinder,S.D.Odintsov and I.Shapipo,Effective Quantum 
Gravity,IOP Press (1992).
\end{thebibliography}
\end{document}